\documentclass[11pt,twoside]{article}
\usepackage{galev06}
\usepackage{psfig}
\usepackage{epsf}
\usepackage{graphics}
\usepackage{lscape}
\def\edcomment#1{\iffalse\marginpar{\raggedright\sl#1\/}\else\relax\fi}
\reversemarginpar

\begin{document}
\title{Extragalactic Science with Herschel-SPIRE}
\author{M. J. Griffin}
\affil{School of Physics and Astronomy, Cardiff University, The Parade, 
Cardiff CF24 3AA, UK}
\author{J. J. Bock \textit{Jet Propulsion Laboratory, Pasadena, California, USA}
\\A. Franceschini \textit{University of Padua, Italy}
\\W. K. Gear \textit{Cardiff University, UK}
\\J. Glenn \textit{University of Colorado, USA}
\\S. Madden \textit{CEA, Saclay, France}
\\S. Oliver \textit{University of Sussex, UK}
\\M. Page \textit{MSSL, Surrey, UK}
\\I. Perez-Fournon \textit{IAC, Tenerife, Spain}
\\ M. Rowan-Robinson \textit{Imperial College, London, UK}
\\ L. Vigroux \textit{Institut d'Astrophysique de Paris, France}
\\ G. Wright \textit{Astronomy Technology Centre, Edinburgh, UK}
\\ and the SPIRE extragalactic science team}

\begin{abstract}
SPIRE, the Spectral and Photometric Imaging Receiver, is one 
of three instruments to fly on ESA's  Herschel Space Observatory. 
It contains a three-band imaging photometer operating at 
250, 360 and 520 $\mu$m, and an imaging Fourier Transform 
Spectrometer (FTS) covering 200-670 $\mu$m. It will be used 
for many extragalactic science programmes, a number of which 
will be implemented as Herschel Key Projects.  The SPIRE 
consortium's  Guaranteed Time programme will devote more 
than 1000 hours to Key Projects covering the high-\textit{z} 
universe and local galaxies. It is also expected that substantial 
amounts of Herschel Open Time will be used for further 
extragalactic investigations. The high-\textit{z} part of the 
SPIRE GT programme will focus on blank-field surveys with a 
range of depths and areas optimised to sample the 
luminosity-redshift plane and characterise the bolometric 
luminosity density of the universe at high \textit{z}.
Fields will be selected that are well covered at other 
wavelengths to facilitate source identifications and enable 
detailed studies of the redshifts, spectral energy distributions, 
and infrared properties of detected galaxies. The local galaxies 
programme will include a detailed spectral and photometric 
study of a sample of well resolved nearby galaxies, a survey 
of more than 300 local galaxies designed to provide a statistical 
survey of dust in the nearby universe, and a study of the ISM in 
low-metallicity environments, bridging the gap between the local 
universe and primordial galaxies. 

\end{abstract}

\vspace{-0.5cm}
\section{Introduction}

The Herschel Space Observatory (Pilbratt, 2004), scheduled for 
launch in 2008, is the fourth cornerstone mission in ESA's science 
programme. Its key science goals are the detection 
and investigation of galaxies at high redshift, and the study of 
star formation and the interstellar medium in our own and nearby 
galaxies.  Herschel will carry a 3.5-m diameter telescope, passively 
cooled to 80 K, and three science instruments: HIFI, PACS and SPIRE. 
The operational lifetime of the mission will be at least three years, 
and approximately two thirds of the observing time will be available 
to the community as Open Time. In this paper we summarise the 
key design features of the SPIRE instrument, and outline its capabilities 
for extragalactic astronomy, using examples from the SPIRE Consortium's 
Guaranteed Time (GT) programme. 
\section{SPIRE instrument design and capabilities}
SPIRE is designed to exploit the particular advantages of Herschel: 
its large-aperture, cold, (80 K), low-emissivity (a few \%) 
telescope; unrestricted access to the poorly explored 200-700 $\mu$m range; 
and the large amount of high quality observing time. It contains a 
three-band  imaging photometer and an imaging Fourier Transform 
Spectrometer (FTS), both of which use feedhorn-coupled bolometer 
arrays cooled to 0.3 K. The photometer field of view is 4x8 
arcmin. Three bolometer arrays are used for broad-band photometry
($\lambda$/$\Delta$$\lambda$ = 3) in spectral bands centred on 
approximately 250, 360 and 520 $\mu$m, with diffraction-limited
beam widths of approx. 18, 25 and 36" respectively.  The same 
field of view is observed simultaneously in the three bands through 
the use of two  dichroic beam-splitters.  Signal modulation can be provided 
either by SPIRE's two-axis Beam Steering Mirror (point source photometry 
or jiggle-mapping) or by scanning the telescope across the sky 
(scan-mapping). The FTS has a 2.6-arcminute diameter field of view 
observed simultaneously by two bolometer arrays covering 200-325 $\mu$m 
and 315-670 $\mu$m. The FTS spectral resolution is adjustable:  the maximum 
resolution for line spectroscopy is 0.04 cm$^{-1}$, for which 
$\lambda$/$\Delta$$\lambda$ varies between 1200 and 300 over
the 200 - 670 $\mu$m range; and the minimum resolution, appropriate for 
continuum spectrophotometry, is 1 cm$^{-1}$ for which $\lambda$/$\Delta$$\lambda$ 
varies from 50 to 15 between 200 and 670 $\mu$m. A detailed description 
of the instrument and its observing modes is given in Griffin 
\textit{et al.} (2006), and the use of a software simulator of 
the photometer system for the optimisation of extragalactic surveys is 
described by Waskett \textit{et al.} (2006).
 

The expected SPIRE sensitivities are summarised in Tables 1 and 2 (all 
sensitivity limits correspond to 5~$\sigma$; 1 hr). 

\begin{table}[!ht]
\caption{Estimated SPIRE photometer sensitivities}
\smallskip
\begin{center}
{\small
\begin{tabular}{ccccc}
\tableline
\noalign{\smallskip}
Band ($\mu$m) & 250 & 360 & 520 &  \\
\noalign{\smallskip}
\tableline
\noalign{\smallskip}
Point source                       & 3.3   & 3.4  & 3.7 & mJy\\
4 x 4 arcmin jiggle map            & 12    & 14   & 16  & mJy\\
Time to map 1 sq. deg to 3 mJy rms & 1.8   & 2.5  & 3.2 & days\\
\noalign{\smallskip}
\tableline
\end{tabular}
}
\end{center}
\end{table}

\begin{table}[!ht]
\caption{Estimated SPIRE spectrometer sensitivities}
\smallskip
\begin{center}
{\small
\begin{tabular}{cccccc}
\tableline
\noalign{\smallskip}
Band ($\mu$m)                   & 200-315 & 315-450 & 450-670 & Units & Mode \\
\noalign{\smallskip}
\tableline
\noalign{\smallskip}
Point source                    & 7.7   & 6.9  & 6.9 - 9.6 & $W m^{-2} \times 10^{-17}$ & Line\\
2.6 arcmin. map                 & 23    & 20   & 20 - 29   & $W m^{-2} \times 10^{-17}$ & Line\\
Point source                    & 260   & 230  & 230 - 320 & mJy                        & Continuum\\
2.6 arcmin. map                 & 760   & 680  & 680 - 950 & mJy                        & Continuum\\
\noalign{\smallskip}
\tableline
\end{tabular}
}
\end{center}
\end{table}

Note that when observing a point source, the photometer produces a 
sparsely sampled 4x4 arcmin. map around the source, and likewise, 
the spectrometer provides a sparsely sampled map of a 2.6 arcmin. 
diameter. A further update of the SPIRE sensitivity model is 
currently in progress, and will take into account as-measured 
instrument performance data. Preliminary results of instrument 
tests to date show that the instrument-level performance is 
generally as expected, although further tests and confirmation are 
needed for the spectrometer.  However, it should be noted that, as 
with many cryogenic infrared space instruments, predicted sensitivity 
figures are subject to large uncertainties (at least a factor of two) 
due to uncertainties the instrument performance in flight and, in 
the case of SPIRE, the effective telescope background.

\section{Extragalactic capabilities of SPIRE}

SPIRE will be used in conjunction with the other Herschel instruments, particularly 
PACS, to carry out a number of coordinated observational programmes.  For the 
investigation of high-\textit{z} galaxies, the PACS and SPIRE photometers will 
together provide a multi-band imager covering the FIR-submm peak in the 
cosmic infrared background; and they will be able to carry out surveys over 
much larger areas than have be done from the ground.

For nearby galaxies, ISO was the first satellite capable of doing FIR spectroscopy 
on nearby galaxies, and demonstrated the value of this in determining the nature, 
excitation, composition of the interstellar medium and the influence of AGN 
activity.  Herschel, using all three instruments, will extend this to modest 
redshift and allow vastly more galaxies to be examined with much better angular 
resolution.   It will be capable of multi-band imaging and spectroscopic studies 
of nearby galaxies to map out the global properties of the ISM and the properties 
of gas and dust in a variety of galaxy types and environments, and to make detailed 
investigation of the impact of metallicity on the ISM and the interaction of the 
ISM with central AGN.  

In this section we give some examples of such projects from the SPIRE consortium's 
GT programme.  All of these observations will be implemented as Herschel Key 
Projects - programmes that will (i) exploit unique Herschel capabilities to address 
important scientific issues in a comprehensive manner, (ii) require a large amount 
of observing time to be used in a uniform and coherent fashion, and (iii) produce 
well characterised and uniform datasets of high archival value.

\subsection{High-redshift galaxies}

Approximately 850 hours of SPIRE GT will be devoted to the high-\textit{z} galaxy 
programme, and the PACS consortium will use about 650 hrs of their GT on a related 
and closely coordinated programme.  This will be one of the flagship projects for 
Herschel and will address important scientific issues such as number count models, 
bolometric (as opposed to single-band) luminosity functions, 
formation and evolution of galaxy bulges and ellipticals, structure formation, 
cluster evolution, the history of energy production, the AGN-starburst connection, 
and cosmic infrared background fluctuations.

The main science driver for the SPIRE high-\textit{z} GT programme is to measure the 
bolometric luminosity density of the Universe as a function of redshift. In 
order to do this it is essential to measure the SEDs of the individual 
sources and to carry out a number of surveys of different depths.  The programme 
therefore consists mainly of a set of 
blank field imaging surveys (to be done in scan-map mode), forming a multi-tiered 
``wedding cake'' covering a range of field sizes, from 0.04 to several tens of 
square degrees, and depths, from a few mJy to several tens of mJy rms. The smaller 
fields will be observed to a 5-$\sigma$ depth comparable to or below the SPIRE 
extragalactic confusion limit (expected to be in the range 20-30 mJy at 40 beams/source) 
depending on the wavelength and the adopted source count model (e.g., 
Vaccari \textit{et al.} 2006).   

The ``wedding cake'' will be designed to sample the luminosity-redshift plane and 
characterise the bolometric luminosity density of the universe at high redshift.  
The wider fields will also sample a range of environments, allowing us to test 
theories of structure formation. Fields will be selected that are well covered 
by XMM-Newton, Spitzer, SCUBA-2, 
PACS-GT and near-IR surveys, to facilitate source identifications and enable 
detailed studies of the redshifts, spectral energy distributions, and infrared 
properties of detected galaxies. 

Some of the deeper wedding cake fields will be used to carry out a multi-band
\textit{P}(\textit{D}) analysis on an area of approx. 1 sq. deg. to measure fluctuations 
down to 3 mJy (about 1 source/beam) and so probe the properties of the 
number counts below the confusion limit.  The availability of multi-band data 
will provide additional diagnostic capabilities in discrimination between 
different source population models.  The survey will also allow a unique search for 
FIR background fluctuations originating from sources below the confusion limit 
and associated with large-scale structure and galaxy clustering.  The shallow 
tiers of the GT survey will be used to investigate clustering on angular 
scales $< 10$ arcmin. (finer spatial scale than can be probed with Planck). 
The background fluctuations on this scale are sensitive to the non-linear 
clustering within a dark matter halo, and the physics underlying the formation 
of far-infrared galaxies within a halo (Cooray \& Sheth, 2002). The large-area 
and multi-wavelength fields will also enable the properties of ordinary 
galaxies below the confusion limit to be investigated through stacking analyses.

The GT programme also includes observations of a sample of 15 rich clusters
between \textit{z} = 0.2 and 1.  Gravitational lensing of background galaxies 
will allow the detection limit to be extended below the blank field confusion 
limit to about 5 mJy. In addition, these observations will be sensitive to
the Sunyaev Zel'Dovich effect, which still produces a significant increment 
in the CMB in the longest-wavelength channel of SPIRE.  The shorter wavelength 
bands will be used to subtract the contribution from cluster galaxies.

Follow-up spectroscopy of selected sources detected in the GT surveys is
expected to be carried out with Herschel and with ground-based facilities 
(ALMA and 10-m class optical/NIR telescopes).  The spectrometers in all 
three Herschel instruments will give unique access to the most important 
cooling lines of interstellar gas, which give very important information 
on the physical processes and energy production mechanisms, and the roles 
of AGN and star formation. 

\subsection{Galaxies in the local universe}

The SPIRE local galaxies GT programme comprises three Key Projects,
requiring approximately 100 hours each: \textit{Physical Processes in 
the ISM of Very Nearby Galaxies}, \textit{The ISM in Low Metallicity 
Environments}, and \textit{The Herschel Galaxy Reference Survey}.  The 
first two are joint PACS-SPIRE projects, and the third is SPIRE only.

\textit{ISM in Nearby Galaxies}: Spatially resolved photometry and spectroscopy with 
SPIRE and PACS will be carried out on a sample of 15 nearby well-studied galaxies, 
including examples of early and late type spirals, low mass spirals, edge-on 
spirals, starburst spirals, starburst galaxies, quiescent dwarfs, starburst dwarfs,
Seyferts, and ellipticals.  Additional spectroscopic data will be obtained
with HIFI.  These observations will allow the detailed SEDs and 
dust properties to be determined, and the variation and evolution of chemistry 
and metallicity to be studied (both within a galaxy and across the range of 
galaxy types.

\textit{Low Metallicity Dwarf Galaxies}:  Much progress has been made 
in characterising galaxies at high redshifts; but the objects discovered so far are 
already metal-rich, implying that they already have a history of star formation 
and metal enrichment processes. Although we are not yet able to observe the earlier 
stage in which primordial galaxies are undergoing their initial episodes of star 
formation, we do have access to low metallicity dwarf galaxies in the local universe
that can serve as analogues to the high-\textit{z} building blocks from which 
galaxies are believed to have formed through mergers

A comprehensive programme of PACS and SPIRE photometry will be implemented 
to study a sample of 55 dwarf galaxies, covering a broad metallicity range 
of 1/50 to 1/3 solar. Additionally, 60 to 600 $\mu$m spectroscopy using PACS 
and HIFI will be obtained on selected sources. The observations will 
shed light on the influence of metallicity on the UV radiation field, gas and dust 
properties, and star-forming activity, the effect of the dust properties on the 
heating and cooling processes in the low-metallicity ISM, and on the impact of 
the super star clusters prevalent in dwarf galaxies on the surrounding gas and dust.

\textit{The Herschel Galaxy Reference Survey}: SPIRE will be used to carry out 
photometry of sample of 320 local galaxies, constituting a benchmark 
survey of dust in the local Universe, and providing the first accurate 
measurements of the amount of dust both inside and outside galaxies.
The primary sample (155 galaxies) comprises objects with 
K(2MASS) $< 9$ (descendents of early universe luminous objects) and with distances 
between 15 and 25 Mpc (allowing the galaxies to be spatially resolved 
with a single pointing).  A secondary sample of sources with K = 9 - 12 
will extend the mass range.  This survey will also help relate 
present-day galaxies to their high-z ancestors, 
and reveal how dust mass and distribution depend on galaxy type, 
environment, and luminosity.  For example, because the sample encompasses 
all environments from the field to rich clusters, it will enable an 
investigation of the as-yet unknown process that appears to inhibit 
star formation in rich environments (Kaufmann et al., 2004)

\subsection{Dust properties}

Understanding dust properties and their dependence on environment is 
vital for the correct interpretation of FIR and submm. observations 
of galaxies (Jones, 2006).  One of the SPIRE galactic Key Projects, 
\textit{The Evolution of Interstellar Dust}, is very relevant in 
this respect, and will provide important results for the extragalactic 
programmes.  It 
involves systematic photometric and spectral surveys of the ISM covering 
the widest possible range of extinction, illumination, density, history, 
and star forming activity.  It will trace the nature and evolution 
of dust in relation to the physical, dynamical and chemical properties 
of the ISM in different environments: diffuse shock processed dust, cirrus, 
molecular clouds, low excitation PDRs, hot PDRs with HII regions, 
pre-stellar cores, and protostars.  The results will allow study of 
the various processes acting on dust particles (fragmentation, coagulation, 
condensation, evaporation, photo processing) in all ISM environments 
from the most tenuous to the most dense.

\section{Conclusions}

SPIRE, in conjunction with the other Herschel instruments, will 
make major advances in extragalactic astronomy. The Guaranteed Time 
programmes summarised here (which take up more than half of the SPIRE 
GT), serve as examples to illustrate the scientific capabilities 
of the instrument and the mission. It is foreseen that there will
be many more extragalactic programmes in Open Time. 

\begin{acknowledgements}
In addition to the authors of this paper, the SPIRE Extragalactic Science Team 
includes: Asier Abreu, Rick Arendt, Herv\'e Aussel, Tom Babbedge, George Bendo, 
Andrew Blain, Jamie Bock, Alessandro Boselli, Veronique Buat, Jordi Cepa, Pierre 
Chanial, Sarah Church, Dave Clements, Asantha Cooray, Jon Davies, Fred C. Dobbs, 
Darren Dowell, Gianfranco De Zotti, Eli Dwek, Simon Dye, Steve Eales, David Elbaz, 
Erica Ellingson, Mark Frost, Frederic Galliano, Ken Ganga, Bruno Guiderdoni, Mark 
Halpern, Evanthia Hatziminaoglou, George Helou, Kate Isaak, Rob Ivison, Guilaine 
Lagache, Glenn Laurent, Brnuo Maffei, Phil Maloney, Hien Nguyen, Alain Omont, 
Pasquale Panuzzo, Marc Sauvage, Richard Savage, Bernhard Schulz, Douglas Scott, 
Luigi Spinoglio, Jason Stevens, Mattia Vaccari, Ian Waddington, 
Tim Waskett, Christine Wilson, Kevin Xu. 
\end{acknowledgements}

\end{document}